\documentclass[preprint,prd,showpacs,nofootinbib]{revtex4}

\usepackage{doi}
\usepackage{hyperref}
\hypersetup{
  colorlinks=true,        
  linkcolor=blue,         
  citecolor=cyan,         
}

\usepackage{bm}
\usepackage{latexsym}
\usepackage{dcolumn}
\usepackage{amsmath,amsfonts,amssymb}
\usepackage{amsthm}
\usepackage{color}
\usepackage{xcolor}
\usepackage{graphicx,float}
\usepackage{comment}
\usepackage{hyperref}

\usepackage{pdfpages}
\usepackage{epstopdf}
\usepackage[utf8]{inputenc}

\def\be {\begin{equation}}
\def\ee {\end{equation}}
\def\bea {\begin{eqnarray}}
\def\eea {\end{eqnarray}}
\def\bc {\begin{center}}
\def\ec {\end{center}}
\def\bfg {\begin{figure}}
\def\efg {\end{figure}}
\def\bi {\begin{itemize}}
\def\ei {\end{itemize}}

\def\beq{\begin{equation}}
\def\eeq{\end{equation}}
\def\br{\begin{eqnarray}}
\def\er{\end{eqnarray}}
\newcommand{\eel}[1] {\label{#1}\end{equation}}

\newcommand{\bdm}{\begin{displaymath}}
\newcommand{\edm}{\end{displaymath}}

\begin{document}

\title{Regular solutions for black strings and torus-like black holes
}
\author{Kimet Jusufi}
\email{kimet.jusufi@unite.edu.mk}
\affiliation{Physics Department, State University of Tetovo, Ilinden Street nn, 1200, Tetovo, North Macedonia}

\begin{abstract}
    In this letter, we extend the recent black hole solution with spherical symmetry [Phys. Lett. B \textbf{835} (2022), 137546], to a new class of black hole solutions with different topology, namely, a torus-like black hole and a black string in the presence of the cosmological constant. We have used the regularized expressions for the gravitational and the electromagnetic potentials, then we derived the corrections to the energy density and, finally, the black hole solutions are obtained which are regular in the limit $r \to 0$. Interestingly, it is found that the total mass of the black hole is corrected due to the regularized self energy of the electrostatic field. In the special limit, we recover the well known solutions in the literature.
\end{abstract}
\maketitle

\section{Introduction}
According to the classical general relativity, we know that the ultimate fate of a spherically symmetric collapsing star must be a black hole which have singularity at the center.  It is mostly believed that the spacetime singularities to be resolved by a quantum theory of gravity. As of today, however, such a theory is not yet fully known. In the past decades, much effort has been made to obtain different regular black hole solutions (see for example \cite{Bardeen,Ayon-Beato:1998hmi,Haward,Frolov,Simpson:2018tsi}). For a review of some regular black hole solutions see also \cite{Sebastiani:2022wbz}, including a recent work which tries to constrain some of these regular black holes with the Event Horizon Telescope image of Sagittarius A$^*$ \cite{Vagnozzi:2022moj}. The interested reader can also see the following works related to the observational effects of regular black holes \cite{Singh:2022dqs,Kumar:2022fqo,Ghosh:2022gka}, and the reference therein.\\

There are, of course, other interesting ways that can be used in order to obtain regular black holes. For example, we note here the concept of T-duality. According to this concept, the mass spectrum does not change if we exchange winding number $w$ and Kaluza-Klein excitation level $n$, namely we can write $w \to n, \,\,\, R \to \alpha^{'2}/R$ \cite{padma,Nicolini:2019irw,Smailagic,Nicolini:2022rlz}. Using ideas from T-duality, an important step forward was made by Nicolini et al. \cite{Nicolini:2019irw}. They basically found a static and spherically
symmetric metric of a black hole with a stringy effect.  In that seminal paper, it was shown how the stringy corrections resemble the Bardeen type solution \cite{Bardeen} and, importantly, it was shown that such a solution has non-perturbative nature due to the stringy corrections. It was therefore natural to extend such a solution to include electric charge. To do so, one has to use regularized expression for the electromagnetic potential reported in Ref.  \cite{Gaete:2022une}. Namely, such a potential was used to obtain regular black hole solution with electric charge in Ref. \cite{k1}. Among other things, in \cite{k1} the rotating solution was obtained, while it was argued how such a solution resembles the well known Ayon-Beato--Garcia spacetime \cite{Ayon-Beato:1998hmi}. Recently, using regularized potentials in $2+1$ dimensions it was shown that indeed there are BTZ regular solutions. The interested reader can see \cite{k2}. Finally, for black holes in 4 dimensional Einstein-Gauss-Bonnet gravity and Verlinde's emergent gravity see \cite{k3,k4}, as well as entropic corrections to Friedmann equations with stringy effect \cite{k5}. \\

Most of the black hole solutions studied in the literature are characterized by the spherical topology. However other black hole solutions with different topology can be shown to exist, at least - mathematically. We point out one such an interesting solution which is electrically charged, static, it has a torus-like topology, and can be constructed by solving the Einstein-Maxwell equations with negative cosmological constant. Such a solution was indeed found by Huang and Liang \cite{Huang} and is known as the torus-like black hole. The thermodynamical properties of the torus-like black hole have been studied for example in \cite{t1}, see also \cite{t2,t3}. Yet another interesting solution by solving the Einstein-Maxwell equations with a cosmological constant was reported by Lemos and Zanchin \cite{Lemos:1995cm}. Their solution is known as the black string. Black strings have been studied in context of Ho\v{r}ava-Lifshitz gravity \cite{Aliev:2011ir}, in addition black string solutions with quintessence fluid have been investigated in Ref. \cite{Ali:2019mxs}, noncommutative inspired black strings \cite{Singh:2017bwj}, charged rotating dilaton black strings in AdS space \cite{Sheykhi:2008rk}, charged rotating black string in gravitating nonlinear electromagnetic fields \cite{Hendi:2013mka}, and many others. Both these solutions, the torus-like black hole and the black string, have a central singularity in the limit $r\to 0$. In the present work, our aim is to extend these solutions by employing regularized expressions for the gravitational/electromagnetic potentials in order to obtain regular solutions.  \\

This paper is outlined as follows. In Section II, we present the solution for the charged torus-like black hole spacetime. In Section
III, we present the solution for the charged black string spacetime. We comment on our results in Section IV.

\section{Regular and charged torus-like black holes }
Let us start by considering a torus-like black hole which can be described by the spacetime metric \cite{Huang} 
\begin{equation}
    ds^2=-f(r)dt^2+\frac{dr^2}{f(r)}+r^2\left(d\theta^2+d\psi^2\right),
\end{equation}
where $0\leq r < \infty$, $0\leq \theta \leq 2\pi$ and $0\leq \psi \leq 2\pi$. For the case of 2-dimensional space with $t=r$=constant, it has the topology $S^1 \times S^1$. In our setup, we have a cosmological constant, electromagnetic field, and the stringy corrections. The total action is therefore written as
\begin{eqnarray}
S=\int d^4x \sqrt{-g} \left[R-2\Lambda-F_{\mu \nu}F^{\mu \nu}\right]+S_m,
\end{eqnarray}
where the action
\begin{eqnarray}
S_m=\int d^4x \sqrt{-g} \mathcal{L},
\end{eqnarray}
describes the stringy corrections. In general there is expression for the Lagrangian of the stringy corrections, however a simple choice for the Lagrangian density is  $\mathcal{L}\sim -\rho$, where the energy density $\rho(r)$, which is yet to be found. 
Using the regularization factor $r \to \sqrt{r^2+l_0^2}$, where $l_0$  is the zero point length, for the gravitational potential we can write
\begin{equation}
    \Phi(r)=-\frac{k M}{ \sqrt{r^2+l_0^2}},
\end{equation}
where $k$ is some constant of proportionality and for the torus-like black hole it can be set to $k=1/\pi$. Of course, $k$ reflects the different geometry and in \cite{Nicolini:2019irw} it is set to unity for the spherical geometry. Note here that $l_0$ in the last expression for the gravitational potential is not simply put by hand. As was shown by Padmanabhan (see, \cite{padma}), one can employ of Padmanabhan's propagator that encodes stringy effects to calculate static potential \cite{Nicolini:2019irw}, meaning that $l_0$ is build in the theory and it is not simply put by hand. By solving the Poisson's equation in in our case 
\begin{equation}
   \nabla^2 \Phi(r)=4 \pi \rho(r),
\end{equation}
we obtain the energy density due to the quantum corrections as follows
\begin{equation}
    \rho(r)=\frac{3\,M\,l_0^2}{4 \pi^2 (r^2+l_0^2)^{5/2}}.
\end{equation}
This is an interesting result and will be used later on. Basically, the stringy effects are encoded via $l_0$, while one can see that the energy density is not described by a point-like distribution but with a smeared mass distribution. Of course, we obtain the standard result when $l_0$ vanishes. In the spirit of \cite{Gaete:2022une}, we now use the regularized electrostatic potential for our torus-like geometry 
\begin{equation}
    V(r)=-\frac{kQ}{\sqrt{r^2+l_0^2}},
\end{equation}
where, again, for the torus-like black hole we set $k=1/\pi$. For the electromagnetic potential we therefore have
\begin{equation}
A_\mu=(-\frac{kQ}{\sqrt{r^2+l_0^2}},0,0,0).
\end{equation}
which is regular when $r \to 0$. We will assume that the form of the energy-tensor the electromagnetic field is
\begin{equation}
 T_{\mu \nu}^{em}=F_{\mu \sigma}{F_{\nu}}^{\sigma}    -\frac{1}{4}g_{\mu \nu}F_{\rho \sigma}F^{\rho \sigma},
\end{equation}
and for the energy density of the energy-momentum field  components we obtain
\begin{eqnarray}
    {\rho}^{em}(r)&=&-{p_r}^{em}(r)=\frac{ Q^2 r^2}{2 \pi^2 (r^2+l_0^2)^3}\\
    {p_{\theta}}^{em}(r)&=&{p_{\psi}}^{em}(r)=\frac{ Q^2 r^2}{2 \pi^2  (r^2+l_0^2)^3}.
\end{eqnarray}

To find the black hole solution, we proceed as follows.  We can either find the mass profile in torus like coordinates which allows us to find the spacetime geometry, or, we can solve the Einstein field equations. Adopting the first picture, we can compute the mass profile via
\begin{eqnarray}
m(r)=4 \pi^2 \int_0^r [\rho(r')+\rho^{em}(r')] r'^2 dr',
\end{eqnarray}
and the general solution is given in the following form 
\begin{eqnarray}
f(r)=Cr^2-\frac{2 m(r)}{\pi r}.
\end{eqnarray}
In what follows, we shall identify the constant $C$ with the cosmological constant and we define $l^2=-3/\Lambda>0$, then by solving for the mass profile we obtain
\begin{equation}
f(r)=\frac{r^2}{l^2}-\frac{2 M r^2}{\pi (r^2+l_0^2)^{3/2}}+\frac{Q^2 \Delta(r)}{2 \pi (r^2+l_0^2)^2}-\frac{3 Q^2}{2 r \pi l_0}\arctan(\frac{r}{l_0})
\end{equation}
where we have defined $\Delta(r)=5r^2+3l_0^2$. Let us comment for a moment here that $M$ is only a mass parameter and not the true mass of the black hole. One can check that in fact the last equation solved the Einstein field equations 
\begin{eqnarray}
G_{\mu\nu}+\Lambda g_{\mu \nu}=8\pi (T^{em}_{\mu \nu}+T_{\mu \nu}),
\end{eqnarray}
by considering for example the $t-t$ component we get 
\begin{equation}
\frac{rf'(r)+f(r)}{r^2}+\frac{4 Q^2 r^2}{\pi (r^2+l_0^2)^3}+\frac{6\,M\,l_0^2}{\pi (r^2+l_0^2)^{5/2}}+\Lambda=0,
\end{equation}
this equation is satisfied, provided $f(r)$ is given by Eq. (14). We can further check the curvature invariants, in particular for the Ricci scalar we get
\begin{eqnarray}
\lim_{r \to 0}R=-\frac{12}{l^2}+\frac{24 M }{\pi l_0^3},
\end{eqnarray}
and for the Kretschmann scalar we obtain
\begin{eqnarray}
\lim_{r \to 0}K=\frac{24}{l^4}+\frac{96 M^2 }{\pi^2 l_0^6}-\frac{96 M }{\pi l_0^3 l^2}.
\end{eqnarray}
We can therefore conclude that the black hole solution (14) is indeed regular as $r\to 0$. Let is point out an interesting aspect of the solution (14). First, if we consider a series expansion around $l_0$, we get 
\begin{eqnarray}
f(r)=\frac{r^2}{l^2}-\frac{2 \left(M+\frac{3 Q^2 \pi }{8l_0}\right)}{\pi r}+\frac{4 Q^2 }{\pi r^2}+\dots
\end{eqnarray}
and then by compering this to the standard torus-like black hole solution which is given by \cite{Huang}
\begin{eqnarray}
f(r)=\frac{r^2}{l^2}-\frac{2M}{\pi r}+\frac{4 Q^2}{\pi r^2},
\end{eqnarray}
we see that, Eq. (19) necessitates the identification of the total mass with 
\begin{eqnarray}
\mathcal{M}=M+\frac{3 Q^2 \pi}{8\, l_0}.
\end{eqnarray}

In other words, the total mass of the black hole $\mathcal{M}$ (or the ADM mass) contains an additional term due to the regularized self energy of the electrostatic field. The mass parameter $M$ can be identified as the bare mass of the black hole. Finally, we can write down the solution for the torus-like black hole in terms of the total mass 
\begin{equation}
f(r)=\frac{r^2}{l^2}-\frac{2 \mathcal{M} r^2}{\pi (r^2+l_0^2)^{3/2}}+\frac{Q^2\Sigma(r)}{2 \pi (r^2+l_0^2)^2}-\frac{3 Q^2}{2 r \pi l_0}\arctan(\frac{r}{l_0}),
\end{equation}
where 
\begin{eqnarray}
\Sigma(r)=\Delta(r)+\frac{3 r^2}{2 l_0 \sqrt{r^2+l_0^2}}.
\end{eqnarray}
With these modifications, the new regular metric (22), is in principle distinguishable from the singular metric (20). However, having in mind that $l_0$ is a small quantity, for large distances $l_0/r \to 0$, the metrics (22) and (20) are effectively the same. The effect of $l_0$ becomes significant only at small distances.

\section{Regular and charged black strings }

Our next example, is the black string solution  which is a cylindrical spacetime
model, for the 2-dimensional space with $t=r$=constant, it has the topology $R \times S^1$. Toward this goal, we shall use the following metric for the black string \cite{{Lemos:1995cm}}
\begin{equation}
ds^2=-f(r)dt^2+\frac{dr^2}{f(r)}+r^2d\phi^2+\alpha^2r^2 dz^2,
\end{equation}
where $0 \leq r < \infty$, $0 \leq \phi < 2\pi$ and $-\infty <z < \infty$.
As for the gravitational potential, we need to chose the following form here
\begin{equation}
    \Phi(r)=-\frac{2 M}{\alpha \sqrt{r^2+l_0^2}}
\end{equation}
where $\alpha^2=l^{-2}=-\Lambda/3>0$, and with $M$ being the mass per unit length. By solving the Poisson's equation in this case 
\begin{equation}
   \nabla^2 \Phi(r)=4 \pi \rho(r),
\end{equation}
we obtain for the energy density the following result
\begin{equation}
    \rho(r)=\frac{3\,M\,l_0^2}{2 \pi \alpha (r^2+l_0^2)^{5/2}}.
\end{equation}
We can already see that when $l_0\to 0$, it reduces to zero, i.e., point mass distribution. Moreover the electromagnetic potential we take 
\begin{equation}
A_\mu=(-\frac{2 Q}{\alpha \sqrt{r^2+l_0^2}},0,0,0),
\end{equation}
where $Q$ is the linear charge density in the $z$-line. 
In this case, for the energy-momentum tensor components we obtain
\begin{eqnarray}
    {\rho}^{em}(r)&=&-{p_r}^{em}(r)=\frac{2 Q^2 r^2}{\alpha^2 (r^2+l_0^2)^3},\\
    {p_{\phi}}^{em}(r)&=&{p_{z}}^{em}(r)=\frac{2 Q^2 r^2}{\alpha^2 (r^2+l_0^2)^3}.
\end{eqnarray}
It is straightforward to show that from the Einstein field equations we get 
\begin{equation}
\frac{rf'(r)+f(r)}{r^2}+\frac{4 Q^2 r^2}{\alpha^2 (r^2+l_0^2)^3}+\frac{12\,M\,l_0^2}{\alpha (r^2+l_0^2)^{5/2}}-3\alpha^2=0.
\end{equation}
This last solution is satisfied, provided the function $f(r)$ is given by 
\begin{eqnarray}
f(r)&=&\alpha^2 r^2-\frac{4Mr^2}{\alpha (r^2+l_0^2)^{3/2}}+\frac{Q^2\Delta(r)}{2 \alpha^2 (r^2+l_0^2)^2}-\frac{3 Q^2}{2 r \alpha^2 l_0}\arctan(\frac{r}{l_0}),
\end{eqnarray}
where we have defined again $\Delta(r)=5r^2+3l_0^2$. For the Ricci scalar we get
\begin{eqnarray}
\lim_{r \to 0}R=-12 \alpha^2+\frac{48 M }{\alpha l_0^3},
\end{eqnarray}
while for the Kretschmann scalar we obtain
\begin{eqnarray}
\lim_{r \to 0}K=24 \alpha^4+\frac{384 M^2 }{\alpha^2 l_0^6}-\frac{192 \alpha M }{ l_0^3}.
\end{eqnarray}
We clearly see that the scalar invariants are finite in the limit $r\to 0$. As in the case of the torus-like black hole, here $M$ is just a mass parameter and not the total mass mass per unit length of the system. To find the total mass, let us consider a series expansion around $l_0$, the last equation gives 
\begin{eqnarray}
f(r)=\alpha^2 r^2-\frac{4\left(M+\frac{3 Q^2 \pi}{16 r \alpha l_0} \right)}{\alpha r}+\frac{4 Q^2}{ r^2 \alpha^2}+\dots.
\end{eqnarray}

Now by compering this expression with the well known solution for the charged black string \cite{Lemos:1995cm}
\begin{eqnarray}
f(r)=\alpha r^2-\frac{4M}{\alpha r}+\frac{4 Q^2}{r^2 \alpha^2},
\end{eqnarray}
we see that it necessitates the identification of the total mass to be 
\begin{eqnarray}
\mathcal{M}=M+\frac{3 Q^2 \pi}{16 \alpha l_0}.
\end{eqnarray}

Interestingly, we found that the todal mass contains a new term due to the regularized self energy of the electrostatic field. We can express the black string spacetime in terms of the total mass as follows
\begin{eqnarray}
f(r)&=&\alpha^2 r^2-\frac{4\mathcal{M}r^2}{\alpha (r^2+l_0^2)^{3/2}}+\frac{Q^2\Sigma(r)}{2 \alpha^2 (r^2+l_0^2)^2}-\frac{3 Q^2}{2 r \alpha^2 l_0}\arctan(\frac{r}{l_0}),
\end{eqnarray}
where we have defined
\begin{eqnarray}
\Sigma(r)=\Delta(r)+\frac{3 r^2}{2 l_0 \sqrt{r^2+l_0^2}}.
\end{eqnarray}

Thus, we found that the stringy corrected black string  described by Eq. (38) differs from the singular metric given by Eq. (36). Although these corrections are very small when observed from large distances i.e. $l_0/r \to 0$. Finally, one can obtain the rotating counterpart of metric (38) by using the transformations \cite{Lemos:1995cm}
\begin{eqnarray}
t \to \Xi t -a \phi,\,\,\,\phi \to \Xi\phi - a \alpha^2 t,
\end{eqnarray}
where $\Xi=\sqrt{1+ a^2 \alpha^2}$, with $a$ being the rotation parameter. The rotating charged black string metric reads 
\begin{equation}
ds^2=-f(r)(\Xi dt -a d\phi)^2+\frac{dr^2}{f(r)}+r^2 \alpha^4 (a dt-\frac{\Xi}{ \alpha^2} d\phi)^2+\alpha^2 r^2 dz^2.
\end{equation}

\section{Conclusions}
 By means of the zero-point length effect, we used a regularized expressions for the gravitational and electromagnetic potentials and found a class of exact black hole solutions. Our first important result in the present work describes a charged and a regular torus-like black hole with a spacetime geometry which is everywhere regular. To obtain the black hole solution, we found the corresponding energy density due to the stringy corrections and the electromagnetic field. We have shown that, in the limit of a vanishing $l_0\to0$, our result resembles the torus-like black hole reported some years ago in Ref. \cite{Huang}. The second important result in this work, concerns the solution of a charged black string spacetime. In doing so, we used the modified energy density for the black string as well as the contribution to the energy density coming from the electromagnetic field. Again, the solution is regular while in the special limit of vanishing $l_0$, our solution resembles the solution reported in Ref. \cite{Lemos:1995cm}. Another important result is that, the total mass of the torus-like and black string black holes are modified due to the regularized self energy of the electrostatic field.  To this end, we have also presented  the general solutions in terms of the total mass of the system.

 The phenomenological aspect of these solutions are of particular interest. Most of the papers concerning the black hole observations use spherically symmetric black hole solutions. However, the shadow images of black holes can be used to test black holes with different topology as was point out in Ref. \cite{Nampalliwar:2020asd}. By considering a spherical accretion model around the black hole, one can basically distinguish these black holes by computing the effect of topology on the electromagnetic intensity and the shadow images. We plan in the near future to apply such a method to the present solutions. We expect the main role to be played by the mass and the charge, while the effect of the minimal length should be very small in BH observations. Such an effect is important in atomic scales due to the fact that $l_0$ is of the Planck length order. Whether we can have observational effects of $l_0$ in black hole observations is an open question, the main issue of course is the huge uncertainty with the present measurement for supermassive black holes (compared to the atomic observations). But for very small black holes predicted bu some quantum gravity theories, we do expect $l_0$ to play an important role. The gravity waves produced during the ring down phase might provide better results compared to the black hole shadow. Finally, in the near future, we plan also to study the stability and the thermodynamics, as well as the energy conditions of these black hole solutions.



\begin{thebibliography}{}

\bibitem{Bardeen} J. M. Bardeen, 
in Proceedings of International Conference GR5, 1968, Tbilisi, USSR, p. 174.

\bibitem{Ayon-Beato:1998hmi}
E.~Ayon-Beato and A.~Garcia,
Phys. Rev. Lett. \textbf{80} (1998), 5056-5059

\bibitem{Haward} S. A. Hayward, 
Phys. Rev. Lett. 96 (2006) 031103

\bibitem{Frolov}
V.~P.~Frolov,
Phys. Rev. D \textbf{94} (2016) no.10, 104056V

\bibitem{Simpson:2018tsi}
A.~Simpson and M.~Visser,
JCAP \textbf{02} (2019), 042;
E.~Franzin, S.~Liberati, J.~Mazza, A.~Simpson and M.~Visser,
JCAP \textbf{07} (2021), 036


\bibitem{Sebastiani:2022wbz}
L.~Sebastiani and S.~Zerbini,
[arXiv:2206.03814 [gr-qc]].

\bibitem{Vagnozzi:2022moj}
S.~Vagnozzi, R.~Roy, Y.~D.~Tsai, L.~Visinelli, M.~Afrin, A.~Allahyari, P.~Bambhaniya, D.~Dey, S.~G.~Ghosh and P.~S.~Joshi, \textit{et al.}
[arXiv:2205.07787 [gr-qc]].

\bibitem{Singh:2022dqs}
B.~P.~Singh, M.~S.~Ali and S.~G.~Ghosh,
[arXiv:2207.11907 [gr-qc]].

\bibitem{Kumar:2022fqo}
J.~Kumar, S.~U.~Islam and S.~G.~Ghosh,
Astrophys. J. \textbf{938} (2022) no.2, 104

\bibitem{Ghosh:2022gka}
R.~Ghosh, M.~Rahman and A.~K.~Mishra,
[arXiv:2209.12291 [gr-qc]].

\bibitem {padma} T. Padmanabhan, Phys. Rev. Lett. 78, 1854 (1997), arXiv:hep- th/9608182.


\bibitem{Nicolini:2019irw}
P.~Nicolini, E.~Spallucci and M.~F.~Wondrak,
Phys. Lett. B \textbf{797} (2019), 134888

\bibitem{Smailagic}
A.~Smailagic, E.~Spallucci and T.~Padmanabhan,
[arXiv:hep-th/0308122 [hep-th]].

\bibitem{Nicolini:2022rlz}
P.~Nicolini,
Gen. Rel. Grav. \textbf{54} (2022) no.9, 106

\bibitem{Gaete:2022une}
P.~Gaete and P.~Nicolini,
Phys. Lett. B \textbf{829} (2022), 137100


\bibitem {k1} P. Gaete, K. Jusufi, and P. Nicolini, Phys. Lett. B \textbf{835} (2022), 137546


\bibitem{k2}
K.~Jusufi,
[arXiv:2209.04433 [gr-qc]].


\bibitem{k3}
K.~Jusufi,
[arXiv:2206.01189 [gr-qc]]

\bibitem{k4}
K.~Jusufi, Annals of Physics,
448 (2023), 169191,

\bibitem{k5}
K.~Jusufi and A.~Sheykhi,
[arXiv:2210.01584 [gr-qc]].


\bibitem{Huang}
C. G. Huang and C. B. Liang,
Physics Letters A,
201, 1, 1995, pages 27-32

\bibitem{t1}
H.~Feng, Y.~Huang, W.~Hong and J.~Tao,
Commun. Theor. Phys. \textbf{73} (2021) no.4, 045403

\bibitem{t2}W. Hong, B. Mu and J. Tao, Int. J. Mod. Phys. D 29 (2020) no.12, 2050078.

\bibitem{t3}Y. W. Han, X. X. Zeng and Y. Hong, Eur. Phys. J. C 79 (2019) no.3, 252.

\bibitem{Lemos:1995cm}
J.~P.~S.~Lemos and V.~T.~Zanchin,
Phys. Rev. D \textbf{54} (1996), 3840-3853; J. P. S. Lemos, Phys. Rev. D 59, 044020 (1999);  J. P. S. Lemos, Class. Quant. Grav. 12, 1081 (1995).

\bibitem{Aliev:2011ir}
A.~N.~Aliev and C.~Senturk,
Phys. Rev. D \textbf{84} (2011), 044010

\bibitem{Ali:2019mxs}
M.~S.~Ali, F.~Ahmed and S.~G.~Ghosh,
Annals Phys. \textbf{412} (2020), 168024

\bibitem{Singh:2017bwj}
D.~V.~Singh, M.~S.~Ali and S.~G.~Ghosh,
Int. J. Mod. Phys. D \textbf{27} (2018) no.12, 1850108

\bibitem{Sheykhi:2008rk}
A.~Sheykhi,
Phys. Rev. D \textbf{78} (2008), 064055

\bibitem{Hendi:2013mka}
S.~H.~Hendi and A.~Sheykhi,
Phys. Rev. D \textbf{88} (2013) no.4, 044044

\bibitem{Nampalliwar:2020asd}
S.~Nampalliwar, A.~G.~Suvorov and K.~D.~Kokkotas,
Phys. Rev. D \textbf{102} (2020) no.10, 104035


\end{thebibliography}
\end{document}